\documentclass[fleqn,12pt]{wlscirep}
\usepackage[utf8]{inputenc}
\usepackage[T1]{fontenc}
\usepackage{url,xcolor}
\usepackage[utf8]{inputenc} 
\usepackage{geometry} 
\usepackage{graphicx}
\usepackage{amssymb}
\usepackage[nolist]{acronym} 
\usepackage{amsmath}
\usepackage{amsfonts, amssymb}
\usepackage{mathtools}
\usepackage{cite}
\usepackage{booktabs} 
\usepackage{upgreek}
\usepackage{bbm}
\usepackage{tikz}

\usepackage{subfig}
\newcommand{\bd}{\begin{description}}
	\newcommand{\ed}{\end{description}}
\newcommand{\be}{\begin{enumerate}}
	\newcommand{\ee}{\end{enumerate}}
\newcommand{\bi}{\begin{itemize}}
	\newcommand{\ei}{\end{itemize}}
\newcommand{\bl}{\begin{list}}
	\newcommand{\el}{\end{list}}
\newcommand{\bt}{\begin{tabbing}}
	\newcommand{\et}{\end{tabbing}}

\newcommand{\EX}[1] {{\mathbb{E}}\left\{{#1}\right\}}

\begin{acronym}
	\acro{BP}{belief propagation}
	\acro{LDPC}{low-density parity-check}
	\acro{MAP}{maximum a posteriori probability}
	\acro{r.v.}{random variable}
	\acro{i.i.d.}{independent, identically-distributed}
	\acro{ML}{maximum likelihood}
	\acro{APP}{a posteriori probability}
	\acro{COMP}{combinatorial orthogonal matching pursuit}
	\acro{LLR}{log-likelihood ratio}
	\acro{ROC}{receiver operating characteristic}
	\acro{BCH}{Bose–Chaudhuri–Hocquenghem}
	\acro{SARS-CoV-2}{severe acute respiratory syndrome coronavirus 2}
	\acro{PCR}{polymerase chain reaction}
	\acro{TMA}{transcription mediated amplification}
	\acro{GT}{group testing}
\end{acronym}

\title
{Identification-Detection Group Testing Protocols for COVID-19 at High Prevalence}

\author[1,*]{Marco Chiani}
\affil[1]{DEI, University of Bologna, Italy}

\author[2]{Gianluigi Liva}
\affil{Institute of Communications and Navigation of the German Aerospace Center (DLR), Germany}

\author[1]{Enrico Paolini}
\affil[*]{marco.chiani@unibo.it}

\begin{abstract}
Group testing allows saving chemical reagents, analysis time, and costs, by testing pools of samples instead of individual samples. We introduce a class of group testing protocols with small dilution, suited to operate even at high prevalence ($5\%-10\%$), and maximizing the fraction of samples classified positive/negative within the first round of tests. Precisely,  if  the  tested  group  has exactly one positive sample then the protocols identify it without further individual tests. The protocols also detect the presence of two or more positives in the group, in which case a second round could be applied to identify the positive individuals. With a prevalence of $5\%$ and  maximum dilution 6, with 100 tests we classify 242 individuals, $92\%$ of them in one round and $8\%$ requiring a second individual test. In comparison, the Dorfman’s scheme can test 229 individuals with 100 tests, with a second round for $18.5\%$ of the individuals. 

\smallskip
\noindent \textbf{Group Testing; COVID-19; Pooling.}

\end{abstract}

\begin{document}

\flushbottom
\maketitle
\thispagestyle{empty}

\section*{Introduction}\label{sec:intro}
We consider those situations where it is necessary to check if some individuals are positive with respect to a given disease. With a direct approach, samples taken from the individuals can be tested one by one, with a number of tests equal to the number of individuals under test. In many cases, however, it is possible to pool samples taken from different individuals and test the pool: if the pool is negative then all the corresponding individuals are declared as negative, while if the pool is positive it means that at least one is positive. Several \ac{GT} techniques based on pooling to reduce the number of tests have been proposed, starting from the work by Dorfman \cite{dorfman}. When the disease prevalence is not too large, this brings considerable savings in terms of tests and therefore chemical reagents, analysis time, effort, and costs. Recently, due also to the cost of sophisticated tests like those based on \acl{PCR} or \acl{TMA}, the use of group testing has been advocated to enable mass screening in the context of the  \ac{SARS-CoV-2} pandemic, with experimental campaigns implemented in a few countries \cite{mallapaty2020mathematical}. 
In \emph{adaptive} group testing, the tests are performed in sequence, with pools that are created based on the outcomes of the previous tests \cite{aldridge2019group,sobel}. On the contrary, in non-adaptive group testing all pools are a-priori set, and tests are carried out in parallel. Both approaches have advantages and shortcomings: adaptive strategies can identify the status of individuals with fewer tests. Nevertheless, considering the time required to carry out each test, a pure adaptive strategy may require an excessive amount of time. Non-adaptive schemes require typically more tests to succeed, but they are faster as tests can be performed in parallel. To combine the advantages of both techniques, while mitigating their limitations, it is sometimes preferable to implement a hybrid approach, where a first screening is performed via a non-adaptive testing step, followed by an adaptive (or even individual) one for the population members that are identified as potentially infected. Approaches of this kind, which date back to the original work of Dorfman \cite{dorfman}, enable remarkable savings in the number of tests. Several current  investigations on the use of group testing for \ac{SARS-CoV-2} screening follow this line\cite{Vandenberg2021,yelin2020,Hirotsu2020,Shental2020,LivPaoChi:21,Ghosh2020,Heidarzadeh2021,mutesa2021}. 
In particular, in the context of group testing for \ac{SARS-CoV-2}, the simple Dorfman approach has  been validated by verifying the sensibility of \acl{PCR} tests with respect to the size $n$ of the pools\cite{yelin2020,Hirotsu2020}. Non-adaptive protocols relying on Reed-Solomon error correcting codes to design the pools have been used to target low infection rate regime (e.g., prevalence below $1.3\%$)\cite{Shental2020}. Bayesian approaches to identify the set of infected samples in a non-adaptive group testing approach have also been addressed\cite{LivPaoChi:21}, as well as schemes that exploit a quantitative knowledge on the viral load in the pools \cite{Ghosh2020,Heidarzadeh2021}. 
Other approaches to group testing in the low prevalence regime exploit a geometrical construction of the pools in a non-adaptive setting \cite{mutesa2021}. 

Differently from previous works, we here are not limited to low prevalence. Specifically, in this paper we describe a new class of protocols for group testing where the main objective is to maximize the probability that classification of samples is completed within the first round of tests. If the tested group has exactly one positive sample, then the protocols detect that there is only one positive and identify it without the need for a second round of individual testing. The protocols also detect the presence (without identification) of two or more positives in the group, in which case a second round must be applied to identify the positive individuals. 
These protocols are thus analogous to error control codes able to correct one error and detect two or more errors occurring in a group \cite{RL09}. Due to this capability to directly identify one positive in the first round, this work is specially suited for high prevalence ($5\%-10\%$) scenarios, differently from other methods which address the low prevalence case  \cite{mallapaty2020mathematical,Shental2020,yelin2020,Hirotsu2020,mutesa2021}. Also, due to the problem of dilution which can lead to false negatives, this work is particularly focused on those pool sizes which can be realistically used in a diagnostic laboratory\cite{mallapaty2020mathematical,Hirotsu2020,ben2020,Abid2020,Vandenberg2021,Barak2021}.  We present next the main results of the investigation, based on the probabilistic analysis detailed at the end of the paper.

\section*{Results}

In the following we will refer to \ac{PCR} for the test, but the procedure is general for any possible test.  
With ``prevalence'' we will indicate the  probability that an individual is positive. 
The direct approach to testing consists of performing individual tests, with one \ac{PCR} for each individual sample, to determine if it is positive or negative. The number of tests in this case equals the number of samples  to classify. 

In \acf{GT}, individual samples are grouped (pooling), and the pools are tested: if a pool is negative it is assumed that all individuals participating to that pool are negative. Thus, the number of \ac{PCR} tests can be reduced with respect to individual testing, if the prevalence is not too high. The saving is more marked for low prevalence. 
In this paper we will refer to \ac{GT} with a first round of pooled tests, possibly followed by a second round of  some (hopefully few) individual tests to complete the classification. 
While the advantage is clear, it must be considered that implementing \ac{GT} imposes a reorganization of the testing process, whose impact should not be underestimated. In fact, with \ac{GT} a phase of preparation of the pools is necessary. This phase should be automated to avoid errors in the processing: this is already possible, as machines currently available in many diagnostic laboratories can be suitably reprogrammed for pooling. Also, while individual testing ends in a single round of \ac{PCR}, in the case of  \ac{GT} it is sometimes necessary to carry out a second round of \acp{PCR} for some individuals (thus requiring additional time). If the number of samples to retest is large, managing the second round, where individual samples needing an individual \ac{PCR} must be reexamined, should be automated to avoid errors and contamination. When full automation of the process is not available, it would be preferred to adopt \ac{GT} schemes with a low fraction of samples needing an individual retest. Also, it must be remarked that large pool sizes can lead to a dilution of the viral load affecting the sensitivity of the test, therefore causing false negatives. For this reason, we will concentrate on schemes with limited pool sizes, which justifies our assumption that the false negative rate is negligible.   

The baseline protocol is that originally proposed by Dorfman in 1943, where: individuals are grouped into groups of $n$; one pool is used to analyze all $n$ individuals (dilution $n$); the mother tubes of the $n$ individuals are set aside; the single pool is tested.
If the pool is negative, all $n$ individuals are declared negative, and no other tests are needed.
If, on the contrary, the pool is positive, it is necessary to carry out a second round of individual tests on all $n$ individuals \cite{dorfman}. 

\subsection*{Identification-Detection for Group Testing: the $Pnp$ Protocols}	

\medskip

We propose a new class of pooling schemes with small dilution, for high prevalence testing scenarios.  Assume a group test employing $p$ pools $P_{1}, P_{2}, \ldots, P_{p}$ to test a group of $n>p$ individuals $I_{1}, I_{2},$  $\ldots, I_{n}$. 
The pooling can be described by a test matrix, where each row is a pool and each column is an individual. The matrix elements are $0$ or $1$, where a $1$ in row $i$ and column $j$ indicates that individual $I_{j}$ participates in pool $P_{i}$.

For identification-detection we propose to use test matrices composed by columns all with a fixed number $c$ of $1$s, so that each individual sample is copied into exactly $c$ pools. With this choice, there would be $c$ positive pools if and only if the group has exactly one positive individual. 
A number of positive pools larger than $c$ indicates that there are two or more positive individuals in the group.  
The largest group size $n$ for a given number of copies $c$ and pools $p$ is 
\begin{eqnarray}\label{eq:npc}
n = \binom{p}{c} \,.
\end{eqnarray} 
We will assume always the largest $n$, as for \ac{GT} the objective is to test the largest possible number of individuals for a given $p$. The pooling matrix columns are thus all possible vectors with $p-c$ elements to $0$ and $c$ elements to $1$. 
The number of individuals per pool (dilution) is indicated as $d$. It can be checked that the dilution for $p$ pools and $n$ individuals, each participating in $c$ pools, is 
\begin{eqnarray}\label{eq:dpc}
d = \binom{p-1}{c-1} \,.
\end{eqnarray} 
Hence, by testing the $p$ pools (first round), the scheme allows to classify immediately the cases of zero positives per group or one positive per group. Therefore, for up to one positive per group there is no need  for individual tests. The scheme also detects the presence of two or more positives per group, in which case a second round of individual tests is required.  

To keep the dilution as small as possible, we investigate in particular the case $c=2$, where each sample is copied in two pools. With this choice, from \eqref{eq:dpc} the dilution is $d=p-1$. We now explicit the test matrices for dilution up to $d=6$. 

\smallskip
\noindent
{\it Protocol $P64$}
\label{example1}

\begin{figure}[!t]
    \centering
    \begin{minipage}{.5\textwidth}
        \centering
	\[
	\begin{pmatrix}
	1 & 1 & 1 & 0 & 0 & 0 \\
	1 & 0 & 0 & 1 & 1 & 0 \\
	0 & 1 & 0 & 1 & 0 & 1 \\
	0 & 0 & 1 & 0 & 1 & 1 
	\end{pmatrix}
	\]
    \end{minipage}%
    \begin{minipage}{0.5\textwidth}
        \centering
\input{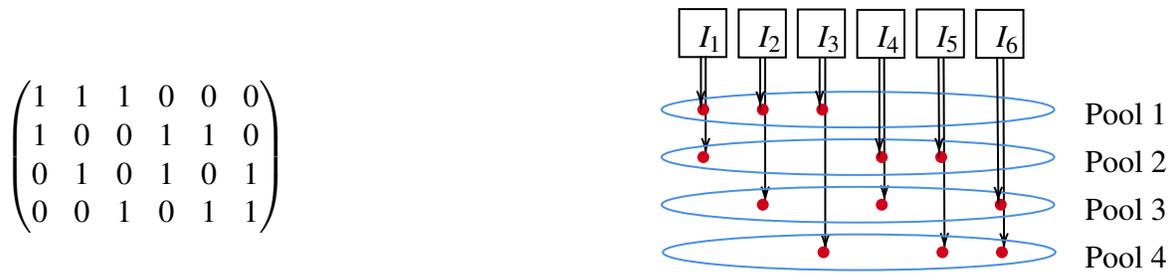}
    \end{minipage}
\caption{Pooling matrix and its interpretation for the $P64$ protocol. $I_{1}, ...,I_{6}$: individuals. }
\label{fig:protocolloP64}
\end{figure}

\noindent From \eqref{eq:npc}, the smallest group size for which $n>p$ is obtained with $p=4$ pools, each individual participating in $c=2$ pools, and therefore with a number of individuals per group $n=6$.
\noindent The test matrix for $P64$ is reported in Fig.~\ref{fig:protocolloP64}. 
Each pool contains the samples from exactly three individuals, so the dilution is $d=3$, as given by \eqref{eq:dpc}. The protocol is described as follows: 
individuals are arranged into groups of $n=6$ (indicated in the figure as $ I_ {1}, ..., I_ {6} $); 
 $p=4$ pools are used to analyze the $6$ individuals;
 each individual participates in $c=2$ pools according to the scheme in the figure, with exactly $3$ individuals in each pool;
the mother tubes of the $6$ individuals are set aside;
the $4$ pools are tested (e.g., by \ac{PCR}).

Based on the results of the $4$ tests, the following cases may arise (see Table \ref{tab:decodifica}):
	\bi
	\item All $4$ pools are negative: in this case all 6 individuals are declared as negative. No other tests are needed.
	\item Exactly $2$ out of the $4$ pools are negative: in this case only one individual is positive, uniquely identified according to the decoding table. No other tests are needed.
	\item One pool is negative and the other $3$ are positive: a second round of individual tests is required for three individuals according to the scheme of Table \ref{tab:decodifica} (or, to simplify, individual test on all $6$ individuals).
	\item All $4$ pools are positive: a second round of individual tests is required for all $6$ individuals.
	\ei

\begin{table}[htb]
\caption{Decision rule for $P64$ ($1=$ positive pool, $0=$ negative pool).}
\centering 
\begin{tabular}{|c|c|c|}
\hline
Pools result & Positive individual & Further test \\
\hline\hline 
0 0 0 0 & None & No \\
1 1 0 0 & $I_{1}$ & No \\
1 0 1 0 & $I_{2}$ & No \\
1 0 0 1 & $I_{3}$ & No \\
0 1 1 0 & $I_{4}$ & No \\
0 1 0 1 & $I_{5}$ & No \\
0 0 1 1 & $I_{6}$ & No \\
0 1 1 1 & $  $ & $I_{4}, I_{5}, I_{6}$ \\
1 0 1 1 & $  $ & $I_{2}, I_{3}, I_{6}$ \\
1 1 0 1 & $  $ & $I_{1}, I_{3}, I_{5}$ \\
1 1 1 0 & $  $ & $I_{1}, I_{2}, I_{4}$ \\
1 1 1 1 & $  $ & $I_{1}, I_{2}, I_{3}, I_{4}, I_{5}, I_{6}$ \\
\hline
\end{tabular}
\label{tab:decodifica}
\end{table}%

\smallskip
\noindent
{\it Protocol $P105$}
\label{exampleP105}

\noindent With $p=5$ pools and $c=2$ we have groups of $n=10$ individuals, pooled according to the test matrix 
	\[
	\begin{pmatrix}
	1 & 1 & 1 & 1 & 0 & 0 & 0 & 0 & 0 & 0\\
	1 & 0 & 0 & 0 & 1 & 1 & 1 & 0 & 0 & 0\\
	0 & 1 & 0 & 0 & 1 & 0 & 0 & 1 & 1 & 0\\
	0 & 0 & 1 & 0 & 0 & 1 & 0 & 1 & 0 & 1\\
	0 & 0 & 0 & 1 & 0 & 0 & 1 & 0 & 1 & 1
	\end{pmatrix} \,.
	\]
In this case the dilution is $d=4$.  This matrix identifies one positive and detects two or more positives per group of $n=10$ individuals.  
 
\noindent
{\it Protocol $P156$}
\label{exampleP156}

\noindent With $p=6$ pools and $c=2$ we have groups of $n=15$ individuals, pooled according to the test matrix 
\[\left(
\begin{array}{ccccccccccccccc}
 1 & 1 & 1 & 1 & 1 & 0 & 0 & 0 & 0 & 0 & 0 & 0 & 0 & 0 & 0 \\
 1 & 0 & 0 & 0 & 0 & 1 & 1 & 1 & 1 & 0 & 0 & 0 & 0 & 0 & 0 \\
 0 & 1 & 0 & 0 & 0 & 1 & 0 & 0 & 0 & 1 & 1 & 1 & 0 & 0 & 0 \\
 0 & 0 & 1 & 0 & 0 & 0 & 1 & 0 & 0 & 1 & 0 & 0 & 1 & 1 & 0 \\
 0 & 0 & 0 & 1 & 0 & 0 & 0 & 1 & 0 & 0 & 1 & 0 & 1 & 0 & 1 \\
 0 & 0 & 0 & 0 & 1 & 0 & 0 & 0 & 1 & 0 & 0 & 1 & 0 & 1 & 1 \\
\end{array} 
\right) \,.
\]
In this case the dilution is $d=5$.  This matrix identifies one positive and detects two or more positives per group of $n=15$ individuals.  

\smallskip
\noindent
{\it Protocol $P217$}
\label{exampleP217}

\noindent With $p=7$ pools and $c=2$ we have groups of $n=21$ individuals, pooled according to the test matrix 
\[\left(
\begin{array}{ccccccccccccccccccccc}
 1 & 1 & 1 & 1 & 1 & 1 & 0 & 0 & 0 & 0 & 0 & 0 & 0 & 0 & 0 & 0 & 0 & 0 & 0 & 0 & 0 \\
 1 & 0 & 0 & 0 & 0 & 0 & 1 & 1 & 1 & 1 & 1 & 0 & 0 & 0 & 0 & 0 & 0 & 0 & 0 & 0 & 0 \\
 0 & 1 & 0 & 0 & 0 & 0 & 1 & 0 & 0 & 0 & 0 & 1 & 1 & 1 & 1 & 0 & 0 & 0 & 0 & 0 & 0 \\
 0 & 0 & 1 & 0 & 0 & 0 & 0 & 1 & 0 & 0 & 0 & 1 & 0 & 0 & 0 & 1 & 1 & 1 & 0 & 0 & 0 \\
 0 & 0 & 0 & 1 & 0 & 0 & 0 & 0 & 1 & 0 & 0 & 0 & 1 & 0 & 0 & 1 & 0 & 0 & 1 & 1 & 0 \\
 0 & 0 & 0 & 0 & 1 & 0 & 0 & 0 & 0 & 1 & 0 & 0 & 0 & 1 & 0 & 0 & 1 & 0 & 1 & 0 & 1 \\
 0 & 0 & 0 & 0 & 0 & 1 & 0 & 0 & 0 & 0 & 1 & 0 & 0 & 0 & 1 & 0 & 0 & 1 & 0 & 1 & 1 \\
\end{array}
\right) \,.
\]
In this case the dilution is $d=6$.  This matrix identifies one positive and detects two or more positives per group of $n=21$ individuals.  

\medskip
\noindent Other protocols, even for different values of $c$, can be similarly designed. For all protocols, the decoding rule can be reformulated succinctly as follows: in the first round, classify as negative all individuals participating in a negative pool, and test individually the others.

\subsection*{Performance}

\noindent
We present the performance of our protocol compared with the Dorfman's protocol, for dilutions ranging from $d=3$ up to $d=6$. 
Considering both the first round of test on $p$ pools and the occasional second round on some or all individuals, for a generic protocol we define the performance in terms of:
\bi
	\item efficiency, quantified by the average number of individuals tested with $100$ \ac{PCR} tests;
	\item probability that an individual is tested in a second round, indicated as $P_{SR}$.  
\ei
We assume a prevalence $\epsilon$ and independent positivity from individual to individual. 
To calculate efficiency, let us denote with $T_{g}$ the number of tests needed to identify all positives in the group. By indicating with $\EX{}$ the statistical expectation, the average number of tests per individual is $\EX{T_{g}}/n$. 
Therefore, on the average, with $100$ \acp{PCR} we classify a number of individuals equal to:
\begin{align}\label{eq:100nTg}
100 \, \frac{n}{\EX{T_{g}}} \quad \text{[individuals classified with $100$ tests]}\,.
\end{align}
The statistical characterization of $T_{g}$ is provided in the Methods section, and leads to equation \eqref{eq:ETg}. 
About $P_{SR}$, we observe that a second round is needed if the number of positive pools, indicated as $Y$, is greater than $2$. The statistic of $Y$, provided in the Methods section, leads to equation \eqref{eq:pSRgen}.

\begin{figure}[p]	
\centering
	\includegraphics[width=0.7\columnwidth,draft=false,clip]{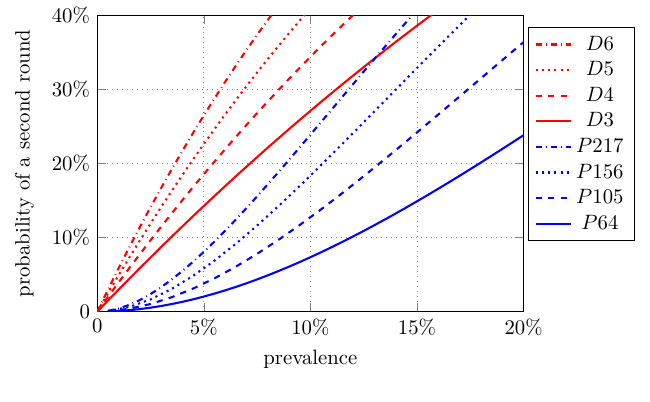}
\caption{Probability of a second testing round for an individual, as a function of the prevalence. $Dn$ = Dorfman with $n$ individuals and one pool; $Pnp$ = protocol P with $n$ individuals and $p$ pools.}
	\label{fig:figPulterioretest}
\end{figure}

\begin{figure}[p]
\centering
	\includegraphics[width=0.7\columnwidth,draft=false,clip]{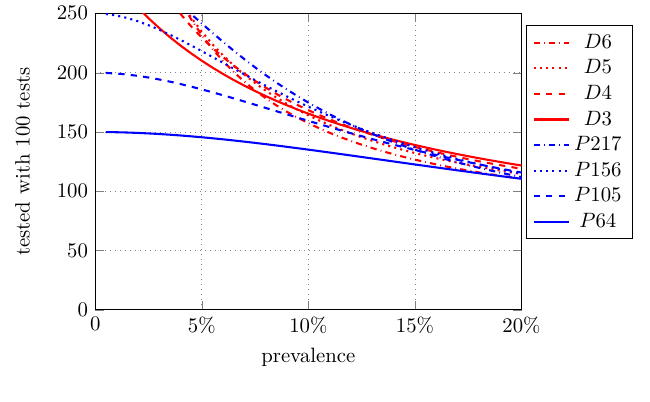}
	\caption{Average number of individuals tested with 100 tests as a function  of the prevalence. $Dn$ = Dorfman with $n$ individuals and one pool; $Pnp$ = protocol P with $n$ individuals and $p$ pools.}
	\label{fig:figTestati_con_100test}
\end{figure}

Results as functions of the prevalence are shown in Fig.~\ref{fig:figPulterioretest} and Fig.~\ref{fig:figTestati_con_100test},  where the probability of a second testing for an individual, given by \eqref{eq:py} and \eqref{eq:pSRgen}, and the efficiency, given by \eqref{eq:100nTg} and \eqref{eq:ETg}, are reported. %
For example, with $P64$ we find that, with a prevalence $\epsilon = 5\% $, about $146$ individuals are classified with $100$ \ac{PCR} tests. Of all individuals, $98\%$ are classified in the first round, and only $2\%$ need a second round.

For the Dorfman's scheme the results, shown in Fig.~\ref{fig:figPulterioretest} and Fig.~\ref{fig:figTestati_con_100test}, have been derived by using \eqref{eq:pSRDorfman} and \eqref{eq:ETgDorfman}. The exact numbers are shown for prevalence of $5\%$ and $10\%$ in Table~\ref{tab:performance}. 
For example, with a pool of $n = 4$ individuals and a prevalence $\epsilon = 5 \%$, an average of $229$ individuals are tested with $100$ \ac{PCR} tests with Dorfman's scheme. However, about $18.5\%$ of all individuals need a second round of individual tests.

\begin{table}[ht]
    \centering
    \begin{tabular}{c|| c |c||c|c||}
         & \multicolumn{2}{c||}{\textbf{prevalence} $5\%$} & \multicolumn{2}{c||}{\textbf{prevalence} $10\%$}\\
         & tested with & retested & tested with & retested \\
         & 100 tests & rate & 100 tests & rate\\ \hline\hline 
        $P64$ & 146 & 2\% & 135 & 7.3\%\\
        $P105$ & 186 & 3.7\% & 159 & 12.7\%\\
        $P156$ & 218 & 5.8\% & 171 & 18.4\%\\
        $P217$ & 242 & 8\% & 175 & 23.9\%\\
        $D3$ & 210 & 14.2\% & 165 & 27.1\%\\
        $D4$ & 229 & 18.5\% & 168 & 34.4\%\\
        $D5$ & 235 & 22.6\% & 164 & 40.9\%\\
        $D6$ & 232 & 26.5\% & 157 & 46.8\% \\
        
    \end{tabular}
    \caption{Performance of the analyzed protocols, dilutions $d=3,\ldots,6$}
    \label{tab:performance}
\end{table}

\section*{Discussion}
We have investigated group testing consisting of a first round of pooled tests, followed by individual testing, applied to a population with high  prevalence, to save resources. 

Two main issues must be discussed for a practical usage of group testing. 
First, one should consider that the maximum pool size is limited due to the dilution of the sample viral load and the consequent problem of false negatives. For COVID-19, a conservative current estimation suggests that dilutions in the order of $5-8$ would still allow a negligible false negative rate, although higher dilutions have been investigated, with some conflicting reports \cite{mallapaty2020mathematical,Hirotsu2020,ben2020,Abid2020,Vandenberg2021,Barak2021}.  
Second, for \acl{GT} the diagnostic laboratory must be organized to handle the whole process (pooling, first round of tests, reopening and second round of tests). Automation systems and robots, currently available in many diagnostic laboratories, can be suitably reprogrammed for pooling. The main issue is related to the management of the second rounds of tests. If the fraction of samples to retest is large, picking back the original samples of some individual to be reexamined should be automated, to avoid errors and contamination. When the process is not fully automated it could be necessary to use protocols able to complete the positive identification mostly within the first round of tests, with a small rate of individuals to be retested. In fact, with low rates of second rounds it may be possible to handle the retesting process even manually, thus  simplifying the organization of a diagnostic laboratory.  Reducing the second round tests will also have the advantage of giving a faster classification. 

Limiting the discussion to dilutions up to $6$, we have found that in the range of prevalence $5\%-10\%$ the best choice is represented by the identification-detection scheme $P217$, which outperforms all the others in terms of efficiency while still having a low rate of second round tests. For example, at $\epsilon=5\%$ it allows to classify $242$ individuals with $100$ tests, with a rate of second round individual tests of about $8\%$. The scheme also performs better than the Dorfman's scheme both in terms of efficiency and rate of second round individual tests. 
Compared with the proposal, the Dorfman's schemes are in fact less efficient and have much larger rates of individuals tested twice (one time in group, then individually). They are therefore not suitable at high prevalence. 
Identification-detection schemes with dilutions $3-5$ give less advantages in terms of number of tests, but offer a smaller rate of second round tests. The scheme $P64$, with dilution $3$, is the one with the smallest rate of retested individuals, having a rate of individual retest of $0.088\%$ at $\epsilon=1\%$, of $2\%$ at $\epsilon=5\%$, and of $7\%$ at $\epsilon=10\%$. 
The choice of the specific protocol imposes therefore, for a given maximum dilution, a trade-off among efficiency and second round rates.

Having to handle few second round tests, all new protocols seems suitable even for non automated laboratories at prevalence up to $5\%$, with few percent of the individuals needing a second round. At prevalence $10\%$ the only protocols with small probability of second round are $P105$ and $P64$, with respectively $12.7\%$ and $7\%$ of the individuals which need retesting. 

We derived also analytical expressions for the performance of the new protocols, allowing the design of identification-detection \acl{GT} pooling schemes for arbitrary dilutions and for the targeted prevalence rates.   

\section*{Methods}

\subsection*{Performance of the $Pnp$ protocols}

\noindent In this section we derive expressions for the performance of the proposed protocols assuming $c=2$, which is the most effective value of $c$ to limit dilution. The analysis can however be generalized to other values of $c$. 

Let us denote as $X$ the number of positive individuals in a group of $n$ individuals, and as $Y$ the number of positive pools out of the $p$ pools. 
For an identification-detection protocol able to identify one single positive in a group of $n$ and detect two or more positives, 
the probability that the group must be reopened for a second round is 
\begin{eqnarray}\label{eq:pGR}
P_{GR}=\Pr\left\{X \geq 2\right\}=\sum_{x=2}^{n}\binom{n}{x} \epsilon^{x}\left(1-\epsilon\right)^{n-x} \,.
\end{eqnarray}
The average number of tests per group can be bounded by assuming that the second round is taken on all $n$ individuals
\begin{eqnarray}\label{eq:ETgbound}
\EX{T_{g}} \leq p+ n \sum_{x=2}^{n}\binom{n}{x} \epsilon^{x}\left(1-\epsilon\right)^{n-x} \,.
\end{eqnarray}
To derive a precise analysis we must consider that the second round occurs on subsets of the group, depending on the number of positive pools. 
To this aim, we observe that the probability that $y$ pools are positive is 
\begin{align}\label{eq:py}
\Pr\left\{Y=y\right\}&= \sum_{x=0}^{n} \Pr\left\{Y=y| X=x\right\} \Pr\left\{X=x\right\} = \sum_{x=0}^{n} a(x,y) \epsilon^{x}\left(1-\epsilon\right)^{n-x} 
\end{align}
where $a(x,y)$ is the number of group configurations with $x$ positive individuals and $y$ positive pools, so that, for example, it is $a(1,2)=n$.  
The values of $a(x,y)$ for arbitrary $x, y$ can be derived by combinatorial analysis. Specifically, we prove at the end of the paper  that $a(x,y)$ is given by the recursion
\begin{align}\label{eq:a_recursive}
a(x,y)=\binom{p}{y} \binom{(y-1) y/2}{x}-\sum _{\ell=1}^{y-1} a(x,\ell) \binom{p-\ell}{p-y} \,.
\end{align}
Values of $a(x,y)$ needed to evaluate $\Pr\left\{Y=y\right\}$ are those for $y>2$, which are reported for some protocols of interest in Tables~\ref{tab:axy64}-\ref{tab:axy217}.  

\begin{table}[h]
\parbox{0.45\textwidth}{\scriptsize 
\centering
\caption{$a(x,y)$ for  $P64$.}
\begin{tabular}{|c|c|c|}
\hline
 $x \diagdown y$ & 3 & 4 \\
\hline\hline 
 2 & 12 & 3 \\
 3 & 4 & 16 \\
 4 & 0 & 15 \\
 5 & 0 & 6 \\
 6 & 0 & 1 \\
\hline
\end{tabular}
\label{tab:axy64}
}
\parbox{0.45\textwidth}{\scriptsize 
\centering
\caption{$a(x,y)$ for  $P105$.}
\begin{tabular}{|c|c|c|c|c|c|}
\hline
 $x \diagdown y$ & 3 & 4 & 5 \\
\hline\hline 
 2 & 30 & 15 & 0 \\
 3 & 10 & 80 & 30 \\
 4 & 0 & 75 & 135 \\
 5 & 0 & 30 & 222 \\
 6 & 0 & 5 & 205 \\
 7 & 0 & 0 & 120 \\
 8 & 0 & 0 & 45 \\
 9 & 0 & 0 & 10 \\
 10 & 0 & 0 & 1 \\
\hline
\end{tabular}
\label{tab:axy105}
}
\end{table}%

\begin{table}[h]
\parbox{0.45\textwidth}{\scriptsize 
\centering
\caption{$a(x,y)$ for  $P156$.}
\begin{tabular}{|c|c|c|c|c|c|}
\hline
 $x \diagdown y$ & 3 & 4 & 5 & 6 \\
\hline\hline 
2 & 60 & 45 & 0 & 0 \\
 3 & 20 & 240 & 180 & 15 \\
 4 & 0 & 225 & 810 & 330 \\
 5 & 0 & 90 & 1332 & 1581 \\
 6 & 0 & 15 & 1230 & 3760 \\
 7 & 0 & 0 & 720 & 5715 \\
 8 & 0 & 0 & 270 & 6165 \\
 9 & 0 & 0 & 60 & 4945 \\
 10 & 0 & 0 & 6 & 2997 \\
 11 & 0 & 0 & 0 & 1365 \\
 12 & 0 & 0 & 0 & 455 \\
 13 & 0 & 0 & 0 & 105 \\
 14 & 0 & 0 & 0 & 15 \\
 15 & 0 & 0 & 0 & 1 \\
 \hline
\end{tabular}
\label{tab:axy156}
}
\parbox{0.45\textwidth}{\scriptsize 
\centering
\caption{$a(x,y)$ for $P217$.}
\begin{tabular}{|c|c|c|c|c|c|}
\hline
 $x \diagdown y$ & 3 & 4 & 5 & 6 & 7 \\
\hline\hline 
 2 & 105 & 105 & 0 & 0 & 0 \\
 3 & 35 & 560 & 630 & 105 & 0 \\
 4 & 0 & 525 & 2835 & 2310 & 315 \\
 5 & 0 & 210 & 4662 & 11067 & 4410 \\
 6 & 0 & 35 & 4305 & 26320 & 23604 \\
 7 & 0 & 0 & 2520 & 40005 & 73755 \\
 8 & 0 & 0 & 945 & 43155 & 159390 \\
 9 & 0 & 0 & 210 & 34615 & 259105 \\
 10 & 0 & 0 & 21 & 20979 & 331716 \\
 11 & 0 & 0 & 0 & 9555 & 343161 \\
 12 & 0 & 0 & 0 & 3185 & 290745 \\
 13 & 0 & 0 & 0 & 735 & 202755 \\
 14 & 0 & 0 & 0 & 105 & 116175 \\
 15 & 0 & 0 & 0 & 7 & 54257 \\
 16 & 0 & 0 & 0 & 0 & 20349 \\
 17 & 0 & 0 & 0 & 0 & 5985 \\
 18 & 0 & 0 & 0 & 0 & 1330 \\
 19 & 0 & 0 & 0 & 0 & 210 \\
 20 & 0 & 0 & 0 & 0 & 21 \\
 21 & 0 & 0 & 0 & 0 & 1 \\ 
 \hline
\end{tabular}
\label{tab:axy217}
}
\end{table}%

Then, we observe that if there are $Y=y$ positive pools the number of individuals to retest in a second round is $\displaystyle \binom{y}{2}$. 
Therefore, the probability that an individual needs a second round is
\begin{eqnarray}\label{eq:pSRgen}
P_{SR}=\frac{1}{n}   \sum_{y=3}^{p} \binom{y}{2}\Pr\left\{Y=y\right\} \,.
\end{eqnarray}
The exact average number of tests needed to classify all $n$ individuals in a group is then 
\begin{eqnarray}\label{eq:ETg}
\EX{T_{g}} = p + \sum_{y=3}^{p} \binom{y}{2}\Pr\left\{Y=y\right\}= p + n  P_{SR} 
\end{eqnarray}
and the number of individuals tested with $100$ tests, defined by \eqref{eq:100nTg}, is rewritten as ${100}/{(p/n +  P_{SR})}$. 
\subsection*{Performance of the Dorfman's scheme}
\noindent
For completeness, we review also the performance for the Dorfman's protocol \cite{dorfman}. In the same hypothesis above, the probability that a second round is needed for the Dorfman's scheme (in this case all individuals have to be retested) is
\begin{eqnarray}\label{eq:pSRDorfman}
P_{SR}=\sum_{x=1}^{n}\binom{n}{x} \epsilon^{x}\left(1-\epsilon\right)^{n-x}=1-\left(1-\epsilon\right)^{n}
\end{eqnarray}
and the average number of tests per group is
\begin{eqnarray}\label{eq:ETgDorfman}
\EX{T_{g}} = 1 + n \, P_{SR} \,.
\end{eqnarray}
The number of individuals tested with $100$ tests is then ${100}/{(1/n +  P_{SR})}$.

\subsection*{Recursive Computation of $a(x,y)$}

\begin{figure}[!h]	
	\includegraphics[width=0.9\columnwidth,draft=false,clip]{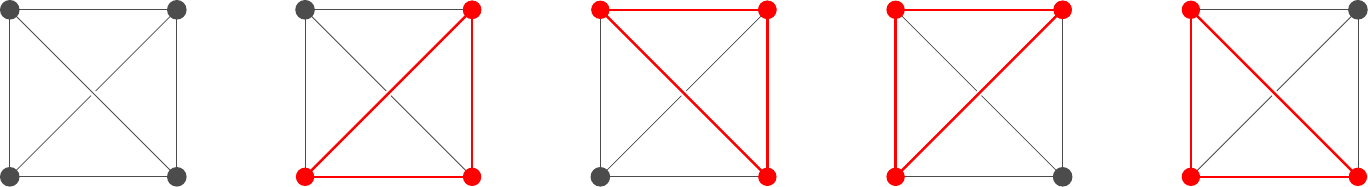}
\caption{Left graph: the $P64$ matrix with four pools (vertices) and six individuals (edges). The four graphs on the right represent the cases with three positive individuals (red edges) producing three positive pools (red vertices).}
	\label{fig:graph}
\end{figure}

When $c=2$, the pooling matrix is amenable of a simple graphical description. In particular, a pooling matrix with $p$ rows and $n$ columns can be represented as a graph $G$ with $p$ vertices, each one associated with a matrix row (equivalent, with a pool), and $n$ edges, each one associated with a matrix column (equivalently, with an individual). An edge connects two vertices if and only if the individual corresponding to the edge participates in the two pools corresponding to the vertices. An example is provided in Fig.~\ref{fig:graph} for the $P64$ pooling matrix.

The value of $a(x,y)$ equals the number of sub-graphs of $G$ having having $y$ vertices (with nonzero degree) and $x$ edges. Since the pooling matrix columns are \emph{all} length-$p$ binary vectors with two $1$s, we can focus on a specific subset $S$ of vertices with cardinality $y$ and search for the number of sub-graphs of $G$ with $y$ vertices (with nonzero degree) and $x$ edges, all vertices belonging to $S$. This number is denoted by $\tilde{a}(x,y)$ and is related to $a(x,y)$ by 
\begin{align}\label{eq:a_tildea_relationship}
    a(x,y) = \binom{p}{y} \tilde{a}(x,y) \,.
\end{align}
The value of $\tilde{a}(x,y)$ equals the number of ways in which, given the set $S$ of $y$ vertices, we can place $x$ edges in such a way that each vertex is connected to at least one edge. This is equal to the total number of ways in which the $x$ edges can be placed, $y(y-1)/2$, minus the number of edge configurations in which only $\ell$ nodes are ``touched'', for $\ell\in\{1,2,\dots,y-1\}$. This yields %
\begin{align}\label{eq:atilde_recursion}
    \tilde{a}(x,y) = \binom{y(y-1)/2}{x} - \sum_{\ell=1}^{y-1} \tilde{a}(x,\ell) \binom{y}{\ell}  
\end{align}
which in particular gives $\tilde{a}(x,2)=1$ if $x=1$ and $\tilde{a}(x,2)=0$ otherwise. 
Incorporating \eqref{eq:atilde_recursion} into \eqref{eq:a_tildea_relationship} gives, after some simplifications, equation \eqref{eq:a_recursive}.

\section*{Acknowledgments}
Te authors would like to thank Prof. Vittorio Sambri for discussions and comments about COVID-19 diagnostic
laboratory activities. Tis work was supported in part by Ministero dell’Istruzione, dell’Università e della Ricerca
(MIUR) under the program "Departments of Excellence (2018-2022) - Precise-CPS".

\end{document}